\documentclass[
 reprint,
 amsmath,amssymb,
 aps,
 nofootinbib,
 superscriptaddress
]{revtex4-2}
\usepackage{graphicx}% Include figure files
\usepackage{bm}% bold math
\usepackage{hyperref}% add hypertext capabilities
\usepackage{booktabs} % nicer tables
\usepackage{physics}
\usepackage{xcolor}
\usepackage{multirow}

\newcommand{\R}[0]{\mathcal{R}}
\newcommand{\PR}[0]{\mathcal{P}_{\R}}
\newcommand{\Ss}[0]{\mathcal{S}}
\newcommand{\picwidth}{0.9\columnwidth}
\newcommand{\picgap}{\hspace{0.8cm}}

\begin{document}

\title{To USR or not to USR: An effective description of two-field dynamics}

\author{Samuel S\'anchez L\'opez}
\email{samuel.lopez@iap.fr}
\affiliation{Institut d’Astrophysique de Paris, 98 bis boulevard Arago, F-75014 Paris, France}
\author{Eemeli Tomberg}
\email{eemeli.tomberg@uclouvain.be}
\affiliation{Cosmology, Universe and Relativity at Louvain (CURL), Institute of Mathematics and Physics, \\ University of Louvain, 2 Chemin du Cyclotron, 1348 Louvain-la-Neuve, Belgium
}

\date{\today}

\begin{abstract}
Single-field models of inflection point inflation experience a phase of ultra-slow roll (USR), leading to amplified curvature perturbations. Recently, Lorenzoni et al. modified the scenario by adding a light spectator field, showing that it ruins USR by setting a floor for the first slow-roll parameter $\epsilon$. Mixing with isocurvature perturbations still amplifies the curvature power spectrum. We offer an alternative point of view, arguing that the inflaton still undergoes effective USR. The spectator lifts the potential by a constant amount, and the inflaton perturbations follow single-field dynamics in the lifted inflection point potential, decoupling from the spectator perturbations. Mode mixing only affects the power spectrum dip, making it milder. We demonstrate this effective single-field behavior numerically for example models. Our results offer a simple way to study the models' parameter dependence.
\end{abstract}

\maketitle

\section{\label{sec:intro}Introduction}
Cosmic inflation produces primordial curvature perturbations, which can seed the structures of the late Universe. On large scales, observations of the cosmic microwave background (CMB) and the large-scale structure constrain the perturbations to be small and to arise from a single degree of freedom \cite{Planck:2018jri}. On smaller scales, no such constraints exist. Models with enhanced small-scale perturbations have attracted much attention for their ability to produce primordial black holes \cite{Hawking:1971ei, Carr:1975qj, Green:2020jor, Carr:2025kdk} and a stochastic background of gravitational waves \cite{Christensen:2018iqi, Domenech:2021ztg}.

Simplest inflationary models are driven by a single scalar field, the inflaton, whose potential determines the perturbation power spectrum. In particular, an inflection point in the potential can enhance the perturbations through a phase called \emph{ultra-slow roll} (USR) \cite{Tsamis:2003px, Dimopoulos:2017ged}. For a review of inflection point models, see \cite{Karam:2022nym}; early examples include \cite{Garcia-Bellido:2017mdw, Kannike:2017bxn, Germani:2017bcs, Ballesteros:2017fsr, Hertzberg:2017dkh, Rasanen:2018fom, Mishra:2019pzq}.

High-energy theories typically contain many scalar fields, and it is natural to expect more than one of them to be active during inflation \cite{Bassett:2005xm}. In this spirit, Refs.~\cite{Lorenzoni:2025gni, Lorenzoni:2025kwn} recently considered adding a free, light spectator field to the inflection point models (see also \cite{Cicoli:2026vab} for an application to axion spectators). The authors showed that even such a minimal extension has a dramatic effect: the model no longer enters USR near the inflection point; instead, the spectator takes over the field velocity, leading to a turning trajectory in field space. Along this trajectory, isocurvature perturbations grow exponentially; as the field turns, they convert into curvature perturbations. The models retain the enhanced curvature power spectrum, but the underlying multi-field mechanism differs drastically from the usual single-field USR.

In this paper, we provide an alternate viewpoint on the models studied in Refs.~\cite{Lorenzoni:2025gni, Lorenzoni:2025kwn}. We argue that the system can still be modeled as single-field, with the inflaton undergoing \emph{effective ultra-slow roll}. The spectator field modifies the inflaton evolution by lifting the potential by an approximately constant amount, but this is the limit of its effect. In particular, if the perturbations are solved in the original field basis instead of the curvature-isocurvature basis, their mixing is minimal for most Fourier modes, and the power spectrum can be reliably solved from the usual single-field equations, taking into account the modified background evolution. We demonstrate the validity and limits of this approach for models lifted from Refs.~\cite{Lorenzoni:2025gni, Lorenzoni:2025kwn}.

The parameters of inflection point models have to be fine-tuned to enhance the power spectrum by a desired amount \cite{Cole:2023wyx, Stamou:2024lqf, Iovino:2025tcv, Profumo:2026qpn}. In Refs.~\cite{Lorenzoni:2025gni, Lorenzoni:2025kwn}, the authors argued that the spectator field alleviates the fine-tuning. This paper makes the spectator's role more transparent, paving the way for a better understanding of this effect.

The paper is organised as follows. In Section~\ref{sec:bg}, we set up the two-field model and discuss background evolution during the ultra-slow-roll-like phase. Section~\ref{sec:perts} compares the perturbation equations in the curvature-isocurvature and field bases, and Section~\ref{sec:numerics} solves them numerically, producing power spectra in different approximations that help estimate the validity of the effective USR picture. We conclude in Section~\ref{sec:conclusions} and relegate technical details to Appendices. Throughout the paper, we use natural units with $M_\text{Pl} = \hbar = c = 1$, where $M_\text{Pl}$ is the reduced Planck mass, $\hbar$ is the reduced Planck constant, and $c$ is the speed of light.

\section{\label{sec:bg}Background evolution of the inflaton-spectator system}

\begin{figure*}
    \centering
    \includegraphics[width=\picwidth]{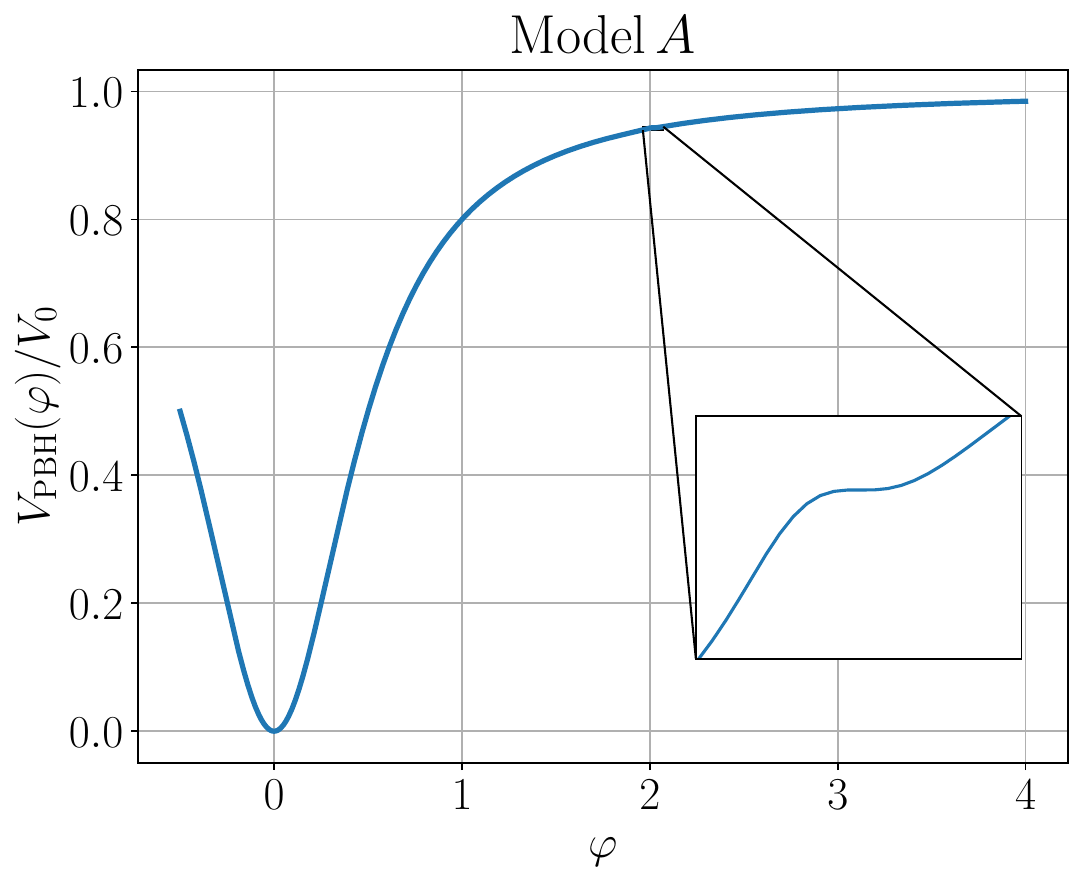}
    \picgap
    \includegraphics[width=\picwidth]{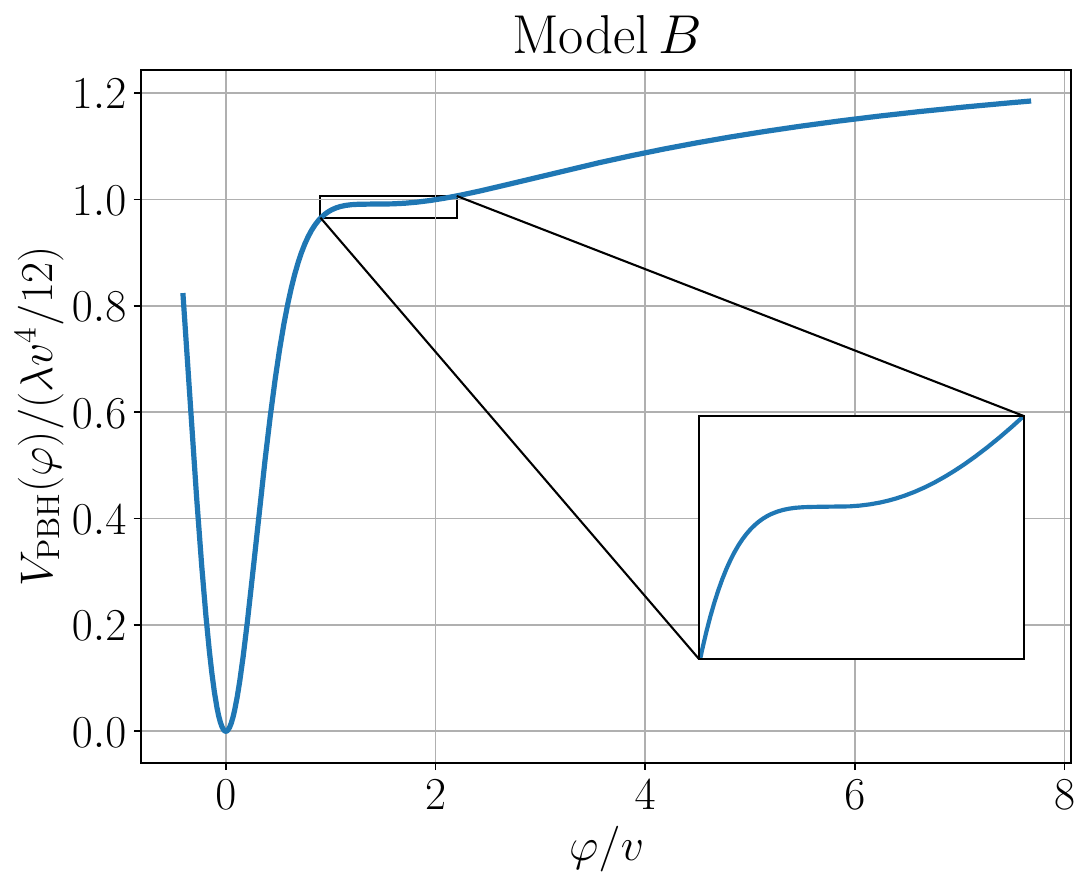}
    \caption{Inflaton potentials \eqref{eq:V_PBH_A} and \eqref{eq:V_PBH_B}. The zoom-ins show the inflection points.}
    \label{fig:potentials}
\end{figure*}

Our inflationary model has two canonical scalar fields: the inflaton $\varphi$, dominating the energy density, and a light spectator $\chi$. The inflationary evolution is determined by the scalar potential. Following Refs.~\cite{Lorenzoni:2025gni, Lorenzoni:2025kwn}, we consider independent potentials for $\chi$ and $\varphi$,
\begin{equation} \label{eq:V}
    V(\varphi, \chi) = V_{\text{PBH}}(\varphi) + V_\text{spect}(\chi)  \, .
\end{equation}
We consider two different scenarios, $A$ and $B$. In both, the spectator is a free field with
\begin{gather}
\label{eq:V_chi}
    V_\text{spect}(\chi) = \frac{1}{2}m_\chi^2\chi^2 \, , \\
\label{eq:chi_mass}
    m_\chi = 6\times 10^{-7} \, (A) \quad \text{or} \quad m_\chi=7\times 10^{-7} \, (B) \, .
\end{gather}
The inflaton potentials read\footnote{Note that Ref.~\cite{Lorenzoni:2025gni} is missing a factor of 2 in the denominator of the exponent in their Eq.~(10). This is corrected in Eq.~(4) of Ref.~\cite{Lorenzoni:2025kwn}.}
\begin{gather}
\label{eq:V_PBH_A}
    A: \quad V_\text{PBH}(\varphi) =
    V_0\frac{\varphi^2}{M^2 + \varphi^2}\qty(1 + A e^{-\frac{(\varphi-\varphi_d)^2}{2\sigma^2}}) \, , \\
\label{eq:A_parameters}
\begin{aligned}
    V_0 &= 8.3\times10^{-11} \, , \quad
    M = \frac{1}{2} \, , \quad
    A = 1.7373\times 10^{-3}\, , \\
    \varphi_d &= 2 (1 - 4\times 10^{-4}) \, , \quad
    \sigma = 1.81\times 10^{-2} \, ,
\end{aligned}
\end{gather}
and
\begin{gather}
\label{eq:V_PBH_B}
    B: \quad V_{\text{PBH}}(\varphi) = \frac{\lambda v^2}{12}
    \frac{\varphi^2\qty(6-4a\varphi/v + 3\varphi^2/v^2)}{\qty(1+b\varphi^2/v^2)^2} \, , \\
\label{eq:B_parameters}
\begin{aligned}
    \lambda &= 1.19\times 10^{-6} \, , &
    v &= 0.19669(1-4\times 10^{-3}) \, , \\
    a &= 0.719527 \, , &
    b &= 1.500016 \, .
\end{aligned}
\end{gather}
They are of the inflection point form, with a local maximum next to a local minimum around $\varphi_d$ in Model $A$ and around
\begin{equation}
    \frac{\varphi}{v} = \frac{b-1 + \sqrt{(b-1)^2 + a^2b}}{ab} \approx 1.73
\end{equation}
in Model $B$; see Fig.~\ref{fig:potentials}.

Model $A$ was first considered in Ref.~\cite{Mishra:2019pzq} and Model $B$ in Ref.~\cite{Garcia-Bellido:2017mdw} (see also Ref.~\cite{Germani:2017bcs}). The chosen parameter values are taken from Ref.~\cite{Lorenzoni:2025kwn} (see Tables I and V therein). Both are compatible with CMB observations.

\subsection{\label{sec:field_evolution}Field evolution}
We collect the inflaton and spectator fields into one vector, $\phi^I = \qty{\varphi,\chi}$. Throughout the paper, we use capital letters $I,J,K$ as indices when referring to field space directions in this basis; when referring to specific components, we replace them with $\varphi$ and $\chi$ (e.g. $\epsilon^\varphi$ and $\epsilon^\chi$ for $\epsilon^I$ in \eqref{eq:directed_sr_parameters_1} below). We use upper and lower indices interchangeably, but so that repeated indices are summed over when (and only when) one of them is up and the other is down.

The fields follow the background equation
\begin{equation} \label{eq:phi_eom}
    \ddot{\phi}^I + 3H\dot{\phi}^I + V^I = 0 \, ,
\end{equation}
where $V^I \equiv \partial_{\phi^I} V$ (similarly for higher derivatives), and the Hubble parameter is given by
\begin{equation} \label{eq:H}
    3H^2 = \frac{1}{2}\dot{\phi}^I\dot{\phi}_I + V \, .
\end{equation}

\begin{figure*}
    \centering
    \includegraphics[width=\picwidth]{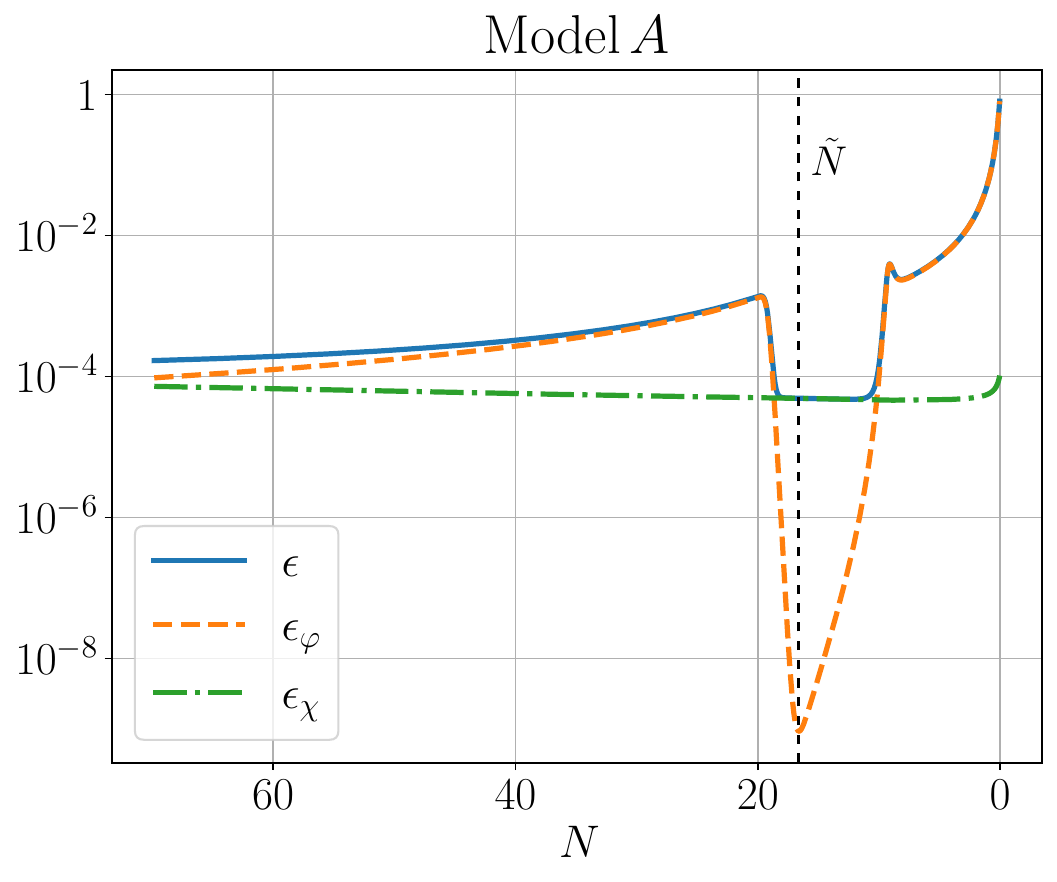}
    \picgap
    \includegraphics[width=\picwidth]{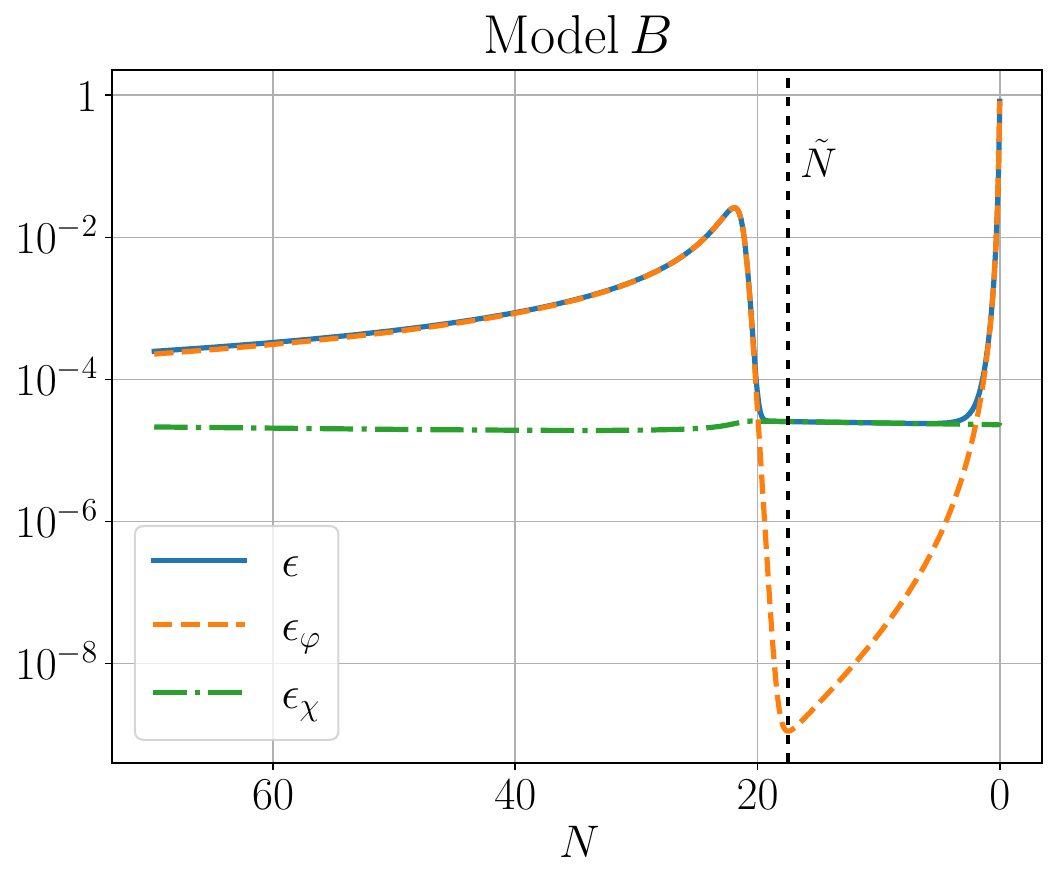}
    \includegraphics[width=\picwidth]{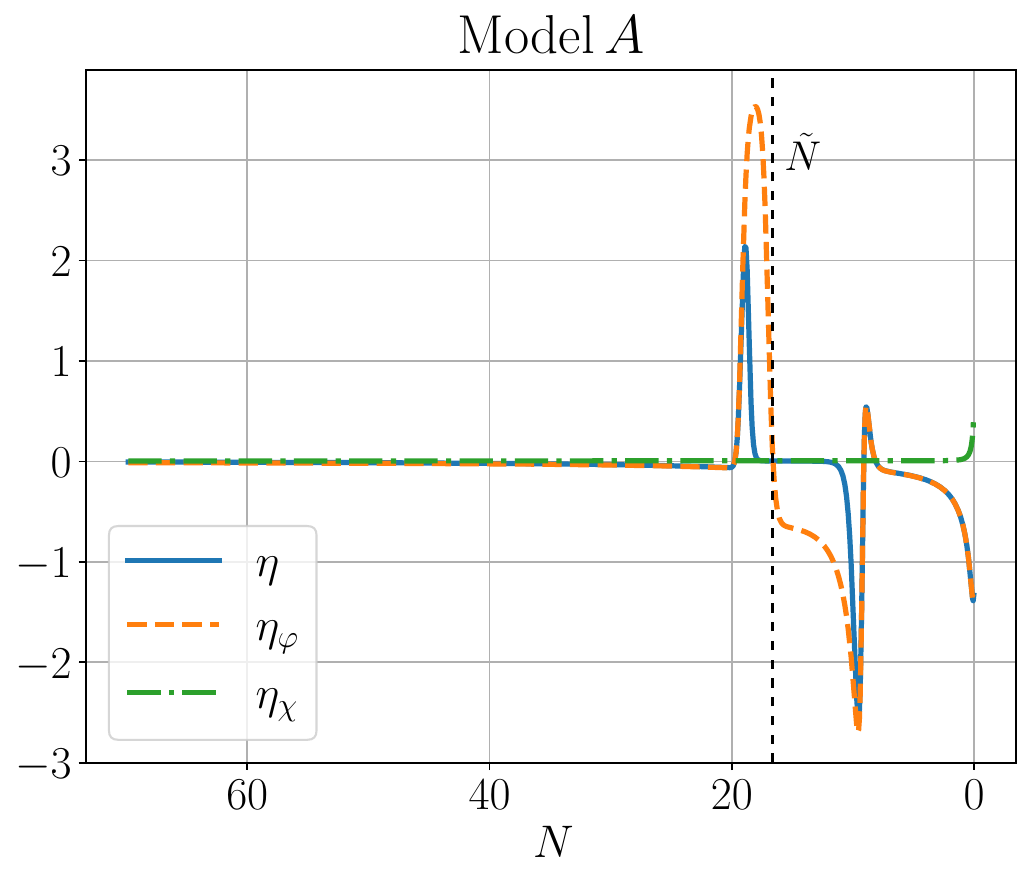}
    \picgap
    \includegraphics[width=\picwidth]{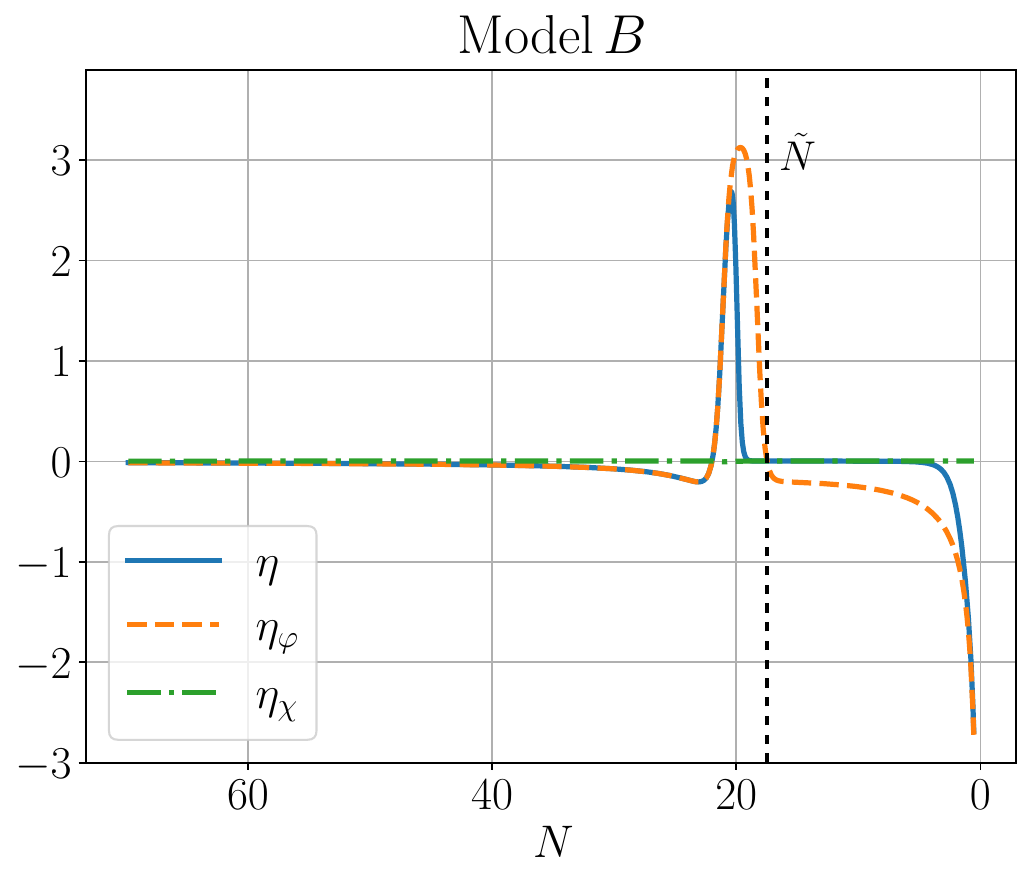}
    \caption{Time evolution of the slow-roll parameters defined in \eqref{eq:sr_parameters}--\eqref{eq:directed_sr_parameters_2} for both models. The e-folds $N$ decrease with time, with $N=0$ at the end of inflation.}
    \label{fig:eps_eta}
\end{figure*}

Following the convention used in Refs.~\cite{Lorenzoni:2025gni, Lorenzoni:2025kwn}, we define the slow-roll parameters
\begin{equation} \label{eq:sr_parameters}
    \epsilon \equiv -\frac{\dot{H}}{H^2} = \frac{\dot{\phi}^I\dot{\phi}_I}{2H^2} \, , \quad
    \delta \equiv \frac{\dot{\epsilon}}{H\epsilon} \, , \quad
    \eta \equiv 2\epsilon - \frac{1}{2}\delta \, .
\end{equation}
We also define the following directed parameters:
\begin{gather}
\label{eq:directed_sr_parameters_1}
    \epsilon^I \equiv \frac{\dot{\phi}^I\dot{\phi}^I}{2H^2} \, , \quad
    \delta^I \equiv \frac{\dot{\epsilon}^I}{H\epsilon^I} \, , \\
\label{eq:directed_sr_parameters_2}
    \eta^I \equiv 2\epsilon - \frac{1}{2}\delta^I = \epsilon - \frac{\ddot\phi^I}{H\dot\phi^I} \, .
\end{gather}
Note that there is no sum over the indices in $\epsilon^I$. We have $\epsilon = \epsilon^\varphi + \epsilon^\chi$; the other parameters don't decompose as neatly, but the directed parameters are useful for the discussion below.

In our models, the potential $V$ does not mix the two fields, so they evolve almost independently of each other, only interacting through the Hubble parameter $H$. For $\chi$ to be a light spectator, we demand $m_\chi^2 \ll H^2$ and $V_\text{spect}(\chi) \ll 3H^2$; it follows that $\chi$ is in slow roll with
\begin{equation} \label{eq:chi_slow_roll}
    3H\dot{\chi} \approx -m_\chi^2\chi \, , \quad
    \frac{1}{2}\dot{\chi}^2 \ll V_\text{spect}(\chi) \, , \quad
    \epsilon_\chi \approx \frac{m_\chi^4\chi^2}{18H^4} \ll 1 \, .
\end{equation}
The inflaton $\varphi$ undergoes the usual behavior of inflection point models: it starts in slow roll, enters an ultra-slow-roll-like phase as it hits the local potential minimum, and transitions to a dual constant roll phase as it rolls over the local potential maximum \cite{Karam:2022nym}. For most of this evolution, the inflaton velocity dominates over that of the spectator, $\epsilon_\varphi > \epsilon_\chi$, but near the maximum, the inflaton velocity momentarily drops to extremely small values and the spectator overtakes it, $\epsilon_\chi \gg \epsilon_\varphi$. The field space trajectory turns, briefly following the $\chi$ direction before aligning with $\varphi$ again.

Fig.~\ref{fig:eps_eta} depicts the evolution of slow-roll parameters as a function of the number of e-folds $N$ before the end of inflation, integrated from $\dd N = -H\dd t$. For future reference, we have denoted the time when $\epsilon_\varphi$ reaches its minimum by $\tilde{N}$.

\subsection{\label{sec:turning_basis}Turning basis}
Instead of the $(\varphi,\chi)$ basis, we can decompose field space vectors in directions parallel and perpendicular to the velocity: the \emph{adiabatic} (curvature) and \emph{entropy} (isocurvature) directions (see, e.g., \cite{Gordon:2000hv, Wands:2007bd, Gong:2016qmq} for reviews). The basis vectors are
\begin{equation} \label{eq:turning_basis}
    \hat\sigma^I \equiv \frac{\dot\phi^I}{\sqrt{\dot\phi^I\dot\phi_I}} \, , \qquad
    \hat s^I \equiv \epsilon^{JI}\hat\sigma_J \, ,
\end{equation}
where $\epsilon^{JI}$ is the Levi-Civita symbol. We use indices $\sigma$ and $s$ to refer to vector components in these directions, e.g. $V_\sigma \equiv V_I \hat \sigma^I$. We denote the turn rate of the velocity by $\omega$, with
\begin{equation} \label{eq:omega}
    \partial_t \hat\sigma^I = \omega \hat s^I \, , \quad
    \partial_t \hat s^I = -\omega \hat \sigma^I \, , \quad
    \omega
    = \frac{V_\varphi \dot{\chi} - V_\chi\dot{\varphi}}{\dot{\varphi}^2 + \dot{\chi}^2} \, .
\end{equation}
For the derivation of \eqref{eq:omega} and further details on the decomposition, see Appendix~\ref{app:bg}.

\subsection{\label{sec:eff_USR}Effective ultra-slow roll}
In single-field inflation, ultra-slow roll refers to evolution where $\eta \geq 3$ (equivalently $\delta \leq -6$), with $\epsilon \ll 1$ and approaching zero fast.\footnote{Originally, ultra-slow roll referred specifically to the case $\epsilon \ll 1$, $\eta = 3$, corresponding to a flat potential $V=\text{const.}$ \cite{Tsamis:2003px, Kinney:2005vj, Dimopoulos:2017ged}. In later literature, the term is also used for $\eta >3$, where the inflaton is climbing up a potential slope \cite{Karam:2022nym}. We adopt the latter convention.} Single-field inflection point models undergo this behavior as the field climbs from the local minimum towards the maximum; the enhancement of the curvature power spectrum can be attributed to this behavior \cite{Karam:2022nym}.

In Refs.~\cite{Lorenzoni:2025gni, Lorenzoni:2025kwn}, the authors argued that the two-field models do not experience USR, since $\epsilon_\chi$ starts to dominate over $\epsilon_\varphi$, quenching the decrease of $\epsilon$ and bringing $\eta$ close to zero. This behavior can indeed be seen in Fig.~\ref{fig:eps_eta}. Instead of conventional USR, the power spectrum enhancement arises from a tachyonic growth of isocurvature perturbations as the field rolls in the $\chi$ direction, converted to curvature perturbations when the field turns back.

While we agree with this observation, we argue in this paper that, instead of the full two-field behavior, it is useful to think of the evolution as a modified single-field system. The inflaton still undergoes essentially the standard USR evolution in the sense that $\epsilon_\varphi$ approaches zero and $\eta_{\varphi} = 3.53 \ (A)$ or $3.12 \ (B)$ near the inflection point, see Fig.~\ref{fig:eps_eta}. The spectator $\chi$ affects the $\varphi$ evolution only through its contribution to $V$ and hence to $H^2$. Since $\chi$ is in slow roll, this contribution is almost constant, shifting the potential energy up uniformly. The enhanced $H$ causes extra friction, bringing the inflaton velocity down; since this velocity is already small, the effect can be significant even for a small change in $H$, prolonging the small-$\epsilon_\varphi$ stage.\footnote{It is crucial that $V$ is here modified additively and not multiplicatively. Multiplying $V$ by a constant factor would simply rescale all time and energy scales, leaving quantities like $\epsilon_\varphi$ intact when computed as a function of a scale-invariant variable such as the number of e-folds of expansion $N = \int \dd t \, H$. Adding a constant term to $V$ can have a non-trivial effect on the evolution. Of course, the contribution from $\chi$ is not strictly constant, but due to its slowly-rolling dynamics, the intuition from the constant case applies.} The relevance of the extra friction was already pointed out in Refs.~\cite{Lorenzoni:2025gni, Lorenzoni:2025kwn, Cicoli:2026vab}.

Below, we will show that the curvature perturbation can be solved essentially equivalently to the single-field case, and that $\epsilon_\varphi$ and $\eta_{\varphi}$ are more important for this solution than the full $\epsilon$ and $\eta$. The second field $\chi$ can be ignored in the perturbation computation, except for its background contribution to $V$. We say the inflaton undergoes a period of \emph{effective ultra-slow roll}, defined as a phase of multi-field inflation during which $\epsilon_\varphi \ll \epsilon \ll 1$ and $\eta_{\varphi} \geq 3$.

\section{\label{sec:perts}Perturbation evolution}

In the spatially flat gauge, the linear field perturbations $Q^I = \qty{\delta \varphi, \delta \chi}$ follow the Fourier space equations of motion (see, e.g., \cite{Gordon:2000hv, Wands:2007bd})
\begin{equation} \label{eq:Q_eom}
\begin{gathered}
    \ddot{Q}_k^I + 3H\dot{Q}_k^I + \frac{k^2}{a^2}Q_k^I + \qty[V^I_J + W^I_J ]Q_k^J = 0 \, , \\
    W^I_J \equiv -\frac{1}{a^3}\frac{\dd}{\dd t}\qty(\frac{a^3}{H}\dot\phi^I\dot\phi_J) \, .
\end{gathered}
\end{equation}
In our setup, $V^I_J$ is diagonal, but $W^I_J$ is not: it mixes the two perturbations.

Instead of $Q^I$, we want to solve the curvature and isocurvature perturbations, related to projections of $Q^I$ onto the turning basis \eqref{eq:turning_basis}:
\begin{equation} \label{eq:R_S}
    \R = \frac{H}{\dot\sigma}Q^I \hat\sigma_I \, , \qquad
    \Ss = \frac{H}{\dot{\sigma}}Q^I \hat s_I \, .
\end{equation}
Using \eqref{eq:Q_eom} and a considerable amount of algebra, one can derive the equations of motion for these quantities:
\begin{gather}
\label{eq:R_eom}
    \frac{\dd}{\dd t}\qty(\dot\R_k - 2\omega\Ss_k) + \qty(3+\delta)H(\dot\R_k-2\omega\Ss_k) + \frac{k^2}{a^2}\R_k = 0 \, , \\
\label{eq:S_eom}
    \ddot\Ss_k + (3+\delta)H\dot\Ss_k + \qty(\frac{k^2}{a^2}+\mu_s^2)\Ss_k = -2\omega\dot\R_k \, ,
\end{gather}
where
\begin{equation} \label{eq:mu}
    \mu_s^2 \equiv V_{ss} - V_{\sigma\sigma} + 2H^2\epsilon\qty(3+\delta-\epsilon)
\end{equation}
with $V_{ss} \equiv V_{IJ}\hat s^I \hat s^J$, $V_{\sigma\sigma} \equiv V_{IJ}\hat \sigma^I \hat \sigma^J$. We outline the derivation of  \eqref{eq:R_eom} and \eqref{eq:S_eom} from \eqref{eq:Q_eom} in Appendix~\ref{app:perts}.

In Refs.~\cite{Lorenzoni:2025gni, Lorenzoni:2025kwn}, the authors analyzed the perturbations in the $(\R$, $\Ss)$ basis using \eqref{eq:R_eom} and \eqref{eq:S_eom}.\footnote{While the discussion of Refs.~\cite{Lorenzoni:2025gni, Lorenzoni:2025kwn} centers around this basis, the authors also used the \texttt{PyTransport} code \cite{Dias:2016rjq, Mulryne:2016mzv}, which solves the correlation functions in the field basis.} In the models at hand, $\mu_s^2 < 0$ during the USR-like phase, since the field is rolling in the $\chi$ direction, so $V_{\sigma\sigma} \approx V_{\chi\chi} = m_\chi^2 > 0$ and $V_{ss} \approx V_{\varphi\varphi} < 0$ near the local maximum, and the last term in \eqref{eq:mu} is $\epsilon$-suppressed. This leads to tachyonic growth in $\Ss_k$. The perturbations mix when the field trajectory turns and $\omega$ is non-zero, and this effect transfers the enhancement to $\R_k$ after the USR-like phase, leading to large curvature perturbations $\R_k$.

\begin{figure*}
    \centering
    \includegraphics[width=\picwidth]{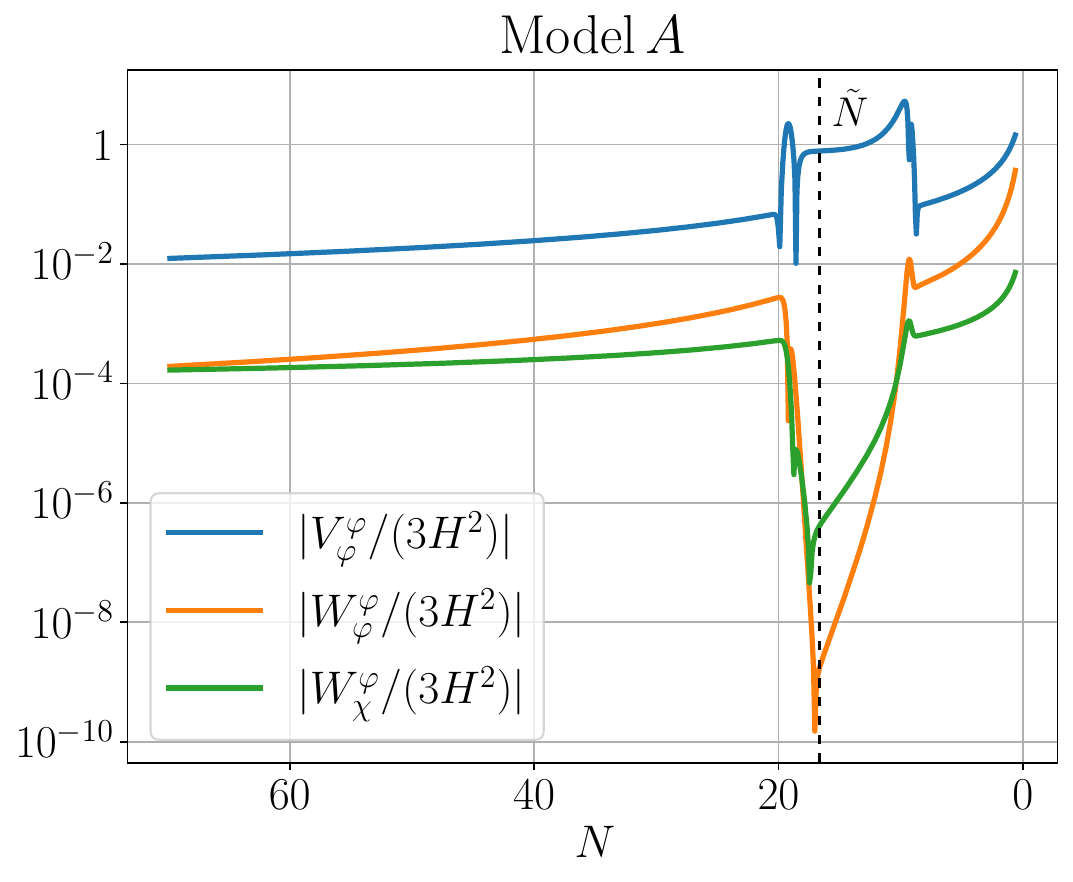}
    \picgap
    \includegraphics[width=\picwidth]{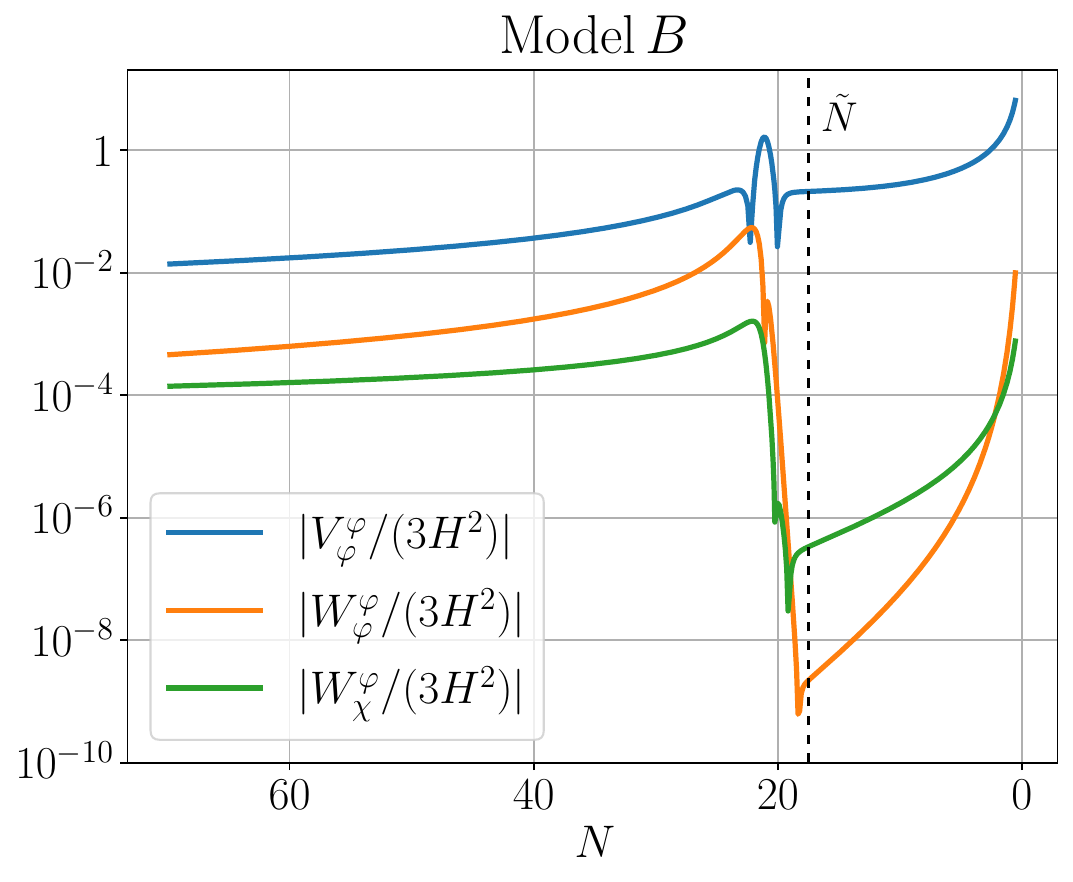}
    \caption{Mass term coefficients in the $Q^\varphi_k$ equation of motion \eqref{eq:Q_eom}.}
    \label{fig:mixing}
\end{figure*}

Instead of the $(\R$,$\Ss)$ basis, we wish to solve the system in the $Q^I$ basis. At late times, the field rolls in the $\varphi$ direction, with $\R \approx H Q^\varphi/\dot\varphi$, so to find $\R$, we're mainly interested in the inflaton perturbations $Q^\varphi$. Just like above, the $V_{\varphi\varphi}$ term in \eqref{eq:Q_eom} is negative during the effective USR phase, leading to tachyonic growth in $Q^\varphi_k$, as already noted in \cite{Lorenzoni:2025kwn}. This term tends to dominate since, as shown in Fig.~\ref{fig:mixing}, the $W^I_J$ elements are completely subdominant compared to $V_{\varphi\varphi}$ during USR, suppressed by the small $\epsilon$. In particular, the $W^\varphi_\chi$ term that would mix $Q^\varphi$ and $Q^\chi$ is negligible, at least if $|Q^\chi| \lesssim |Q^\varphi|$. If we ignore this term, $Q^\varphi$ evolves independently of $Q^\chi$ and identically to a single-field USR model. Instead of the full two-field slow-roll parameters, the evolution of $Q^\varphi$ depends on the directed parameter $\eta_\varphi$ (essentially corresponding to $V_{\varphi\varphi}$ \cite{Karam:2022nym}), and relating it to $\R$ depends on $\epsilon_\varphi$ (through $\dot\varphi$), as we suggested at the end of Section~\ref{sec:bg}.\footnote{We note that in \cite{Lorenzoni:2025kwn}, the authors did briefly study the perturbation evolution from the perspective of the $Q^I$ variables, but they did not consider solving the system with an uncoupled $Q^\varphi$ only.}

In the next section, we will solve the perturbations numerically in the $Q^I$ basis to test the validity of this analysis. We initialize the modes in the Bunch--Davies vacuum state at sub-Hubble scales and follow them until the end of inflation, when $\R$ has frozen to its final super-Hubble value. To solve the quantum system of mixed modes, we actually need to track two sets of mode functions: $Q^\varphi_{1,k}$ which starts from the Bunch--Davies state mixing with $Q^\chi_{1,k}$ which starts from zero, and $Q^\chi_{2,k}$ which starts from its own Bunch--Davies state mixing with $Q^\varphi_{2,k}$ which starts from zero. The technical details of quantization are explained in Appendix~\ref{app:quantization}. The final curvature power spectrum is given by the sum
\begin{equation} \label{eq:PR}
\begin{aligned}
    \PR(k) &= \frac{k^2}{2\pi^2}\sum_{n=1,2} \qty|\frac{H}{\dot\sigma}Q_{n,k}^{I}\hat\sigma_I|^2 \\
    &\approx
    \frac{k^2}{2\pi^2}\sum_{n=1,2} \qty|\frac{H}{\dot\varphi}Q_{n,k}^\varphi|^2 \, .
\end{aligned}
\end{equation}

\section{\label{sec:numerics}Numerical solutions}

We solve numerically the background equations \eqref{eq:phi_eom}--\eqref{eq:H} together with the $Q^I$ basis perturbation equations \eqref{eq:Q_eom} for our two models. We use a non-uniform grid for the comoving wavenumber $k$, with 20 points on large scales and a denser set of 200 points on small scales, in order to properly sample features in the spectrum. We initialize the modes 8 e-folds before horizon crossing and solve their evolution until the end of inflation with both absolute and relative integration tolerances set to $10^{-8}$. We have ensured numerical convergence by varying both the initialization time and tolerances until the resulting spectra were stable.

\begin{figure*}
    \centering
    \includegraphics[width=\picwidth]{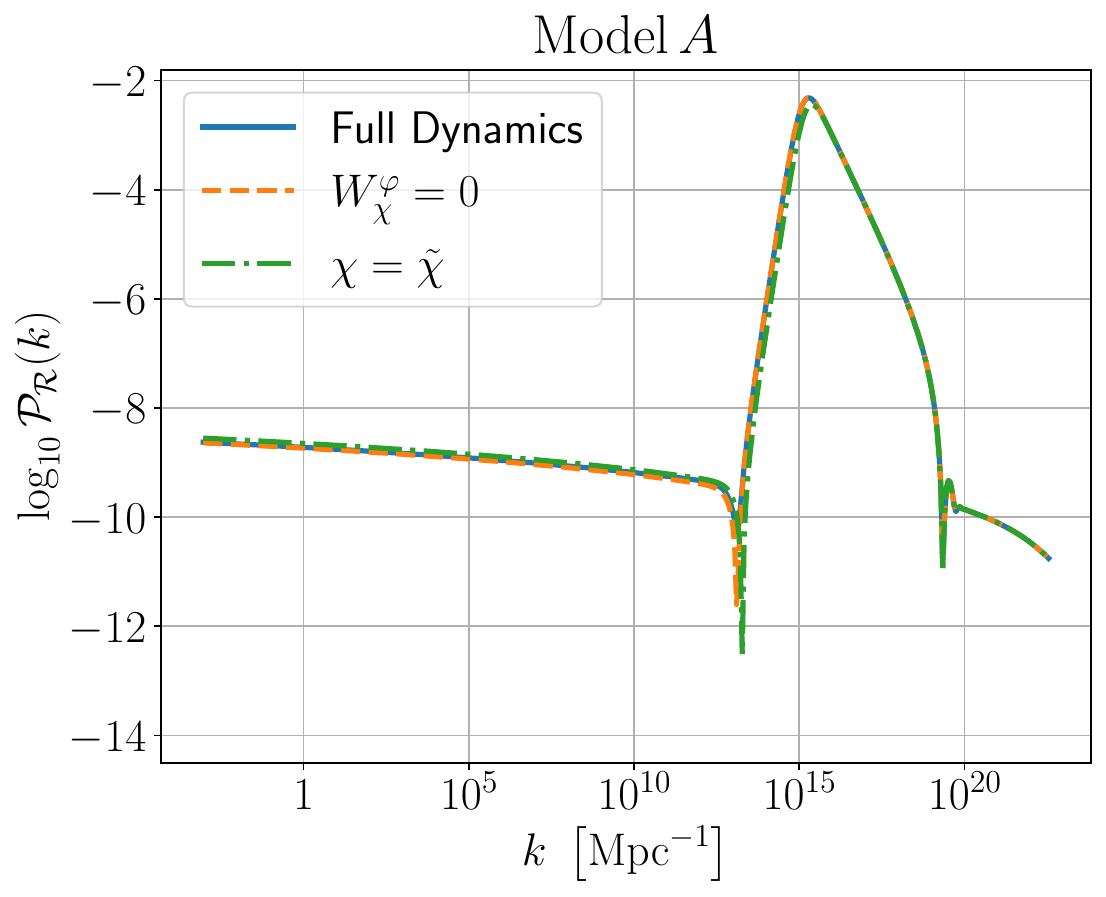}
    \picgap
    \includegraphics[width=\picwidth]{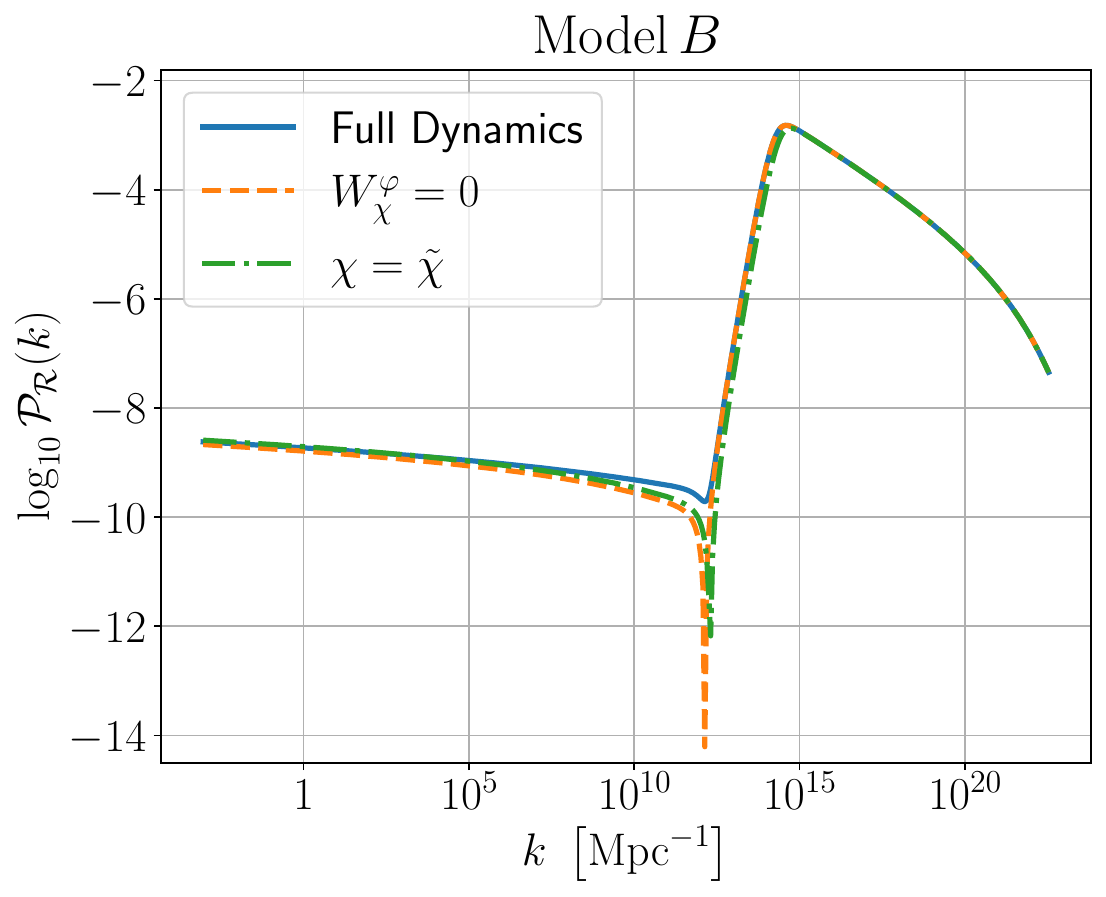}
    \caption{Curvature perturbation power spectra, computed in three ways: with the full dynamics, setting the mixing term $W_\chi^\varphi$ to zero, and replacing the spectator field by a constant energy density $V_\chi(\tilde{\chi})$.}
    \label{fig:PR}
\end{figure*}

The resulting curvature power spectra are shown in blue in Fig.~\ref{fig:PR}. They closely resemble the power spectra of single-field inflection point models, with a CMB plateau at large scales, followed by a narrow dip and a broader peak \cite{Karam:2022nym}. Below, we denote the location of the dip (the power spectrum minimum) by $k_\text{dip}$ and the location of the peak (the power spectrum maximum) by $k_\text{peak}$. The spectra closely match those obtained in \cite{Lorenzoni:2025kwn}. We have also checked that we obtain a matching power spectrum by solving the perturbations in the $(\R,\Ss)$ basis.

To test the role of the spectator field, we have additionally solved the dynamics modified in two different ways: by decoupling the equations of motion for $Q_k^\varphi$ and $Q_k^\chi$, and by replacing the dynamical spectator field with a constant energy density.

For the first modification, we set the off-diagonal terms in the mixing matrix $W^I_J$ to zero, $W^\varphi_\chi = W^\chi_\varphi = 0$, while leaving the background dynamics unchanged. In other words, we solve the modified perturbation equations 
\begin{align}
        &\ddot{Q}_k^\varphi + 3H\dot{Q}_k^\varphi + \frac{k^2}{a^2}Q_k^\varphi + \qty[V^\varphi_\varphi + W^\varphi_\varphi ]Q_k^\varphi = 0 \, , \label{eq:Qphidecoupled}\\
        &\ddot{Q}_k^\chi + 3H\dot{Q}_k^\chi + \frac{k^2}{a^2}Q_k^\chi + \qty[V^\chi_\chi + W^\chi_\chi ]Q_k^\chi = 0 \, .\label{eq:Qchidecoupled}
\end{align}
In this limit, $Q^\varphi_{2,k}=Q^\chi_{1,k}=0$, and only the standard Bunch--Davies mode functions contribute. The approximation lets us test the relevance of the mixing between $Q_k^\varphi$ and $Q_k^\chi$, which we argued above to be small.

For the second modification, we go one step further and replace the spectator field with a constant contribution to the potential energy density, given by 
\begin{equation}
    V_{\rm spect}(\tilde{\chi}) = \frac{1}{2}m_{\chi}^2\tilde{\chi}^2 \, ,
\end{equation}
leading to a modified background
\begin{equation}
    3H^2 = \frac{1}{2}\dot{\varphi}^2 + V_{\rm PBH}(\varphi) + V_\chi(\tilde{\chi}) \, .
\end{equation}
Here $\tilde{\chi}$ is the spectator field value at $N=\tilde{N}$, that is, when $\epsilon_\varphi$ reaches its minimum in the full two-field dynamics. We chose this $\chi$ value because we expect evolution around $\tilde{N}$ to be the most important for the power spectrum peak. The perturbation dynamics reduce to the single-field case,
\begin{align}
        \label{eq:Qphidecoupled_2}
        &\ddot{Q}_k^\varphi + 3H\dot{Q}_k^\varphi + \frac{k^2}{a^2}Q_k^\varphi + \qty[V^\varphi_\varphi + W^\varphi_\varphi ]Q_k^\varphi = 0 \, .
\end{align}
This approximation is strictly less accurate, but also simpler, than the previous one. One may expect it to be good when, in addition to the mode mixing being negligible, also $\epsilon_\chi \ll 1$, so that $\chi$ stays frozen.

The curvature power spectra obtained with these approximations are shown  in Fig.~\ref{fig:PR} in orange and green, respectively. The spectra are in good agreement with the full result around the CMB scales and the peak, while some differences appear around the dip that precedes the peak, especially in Model $B$.

The peak is the most interesting feature. Table~\ref{table:maxima} gives the peak locations $k_\text{peak}$ together with the peak heights, in the different approaches for both models. The no-mixing results $W^\varphi_\chi=0$ are practically identical to the full results: the $k_\text{peak}$ values are identical within our $k$ resolution, and the power spectra are identical to 3 significant digits for Model $A$ and to 
5 significant digits for Model $B$.
The frozen $\chi=\tilde{\chi}$ results have a $k_\text{peak}$ value that is higher than that of the full result by 0.3 e-folds in Model $A$ and 0.4 e-folds in Model $B$, while the power spectra are slightly lower than in the full case, by $25\%$ for Model $A$ and by $11\%$ for Model $B$. Since we're primarily interested in the order of magnitude of $\PR$, the approximation is still good.

\begin{table}
    \centering %RevTex sucks and clashes with the array package, so need to adjust clumns widths artificially with hspace.
    \begin{tabular}{llcc}
        \toprule
         Model\hspace{0.8cm} & Approach\hspace{0.1cm} & $\hspace{0.3cm}\log_{10}\frac{k_\text{peak}}{\mathrm{Mpc}^{-1}}$\hspace{0.3cm} & $\log_{10}\mathcal{P}_{\mathcal{R}}$ \\
        \midrule
        \multirow{3}{*}{Model $A$} &Full Dynamics & 15.26 & -2.317681  \\
        &$W_{\chi}^{\varphi}=0$ & 15.26 & -2.317757  \\
        &$\chi=\tilde{\chi}$ & 15.37 & -2.440017 \\
        \midrule
        \multirow{3}{*}{Model $B$} &Full Dynamics & 14.58 & -2.819442 \\
        &$W_{\chi}^{\varphi}=0$ & 14.58 & -2.819443 \\
        &$\chi=\tilde{\chi}$ & 14.76 & -2.870455 \\
        \bottomrule
    \end{tabular}
        \caption{Peak wavenumber and curvature power spectrum amplitude for the full dynamics and both approximations, for Models $A$ and $B$.}
        \label{table:maxima}
\end{table}

\begin{figure*}
    \centering
    \includegraphics[width=\picwidth]{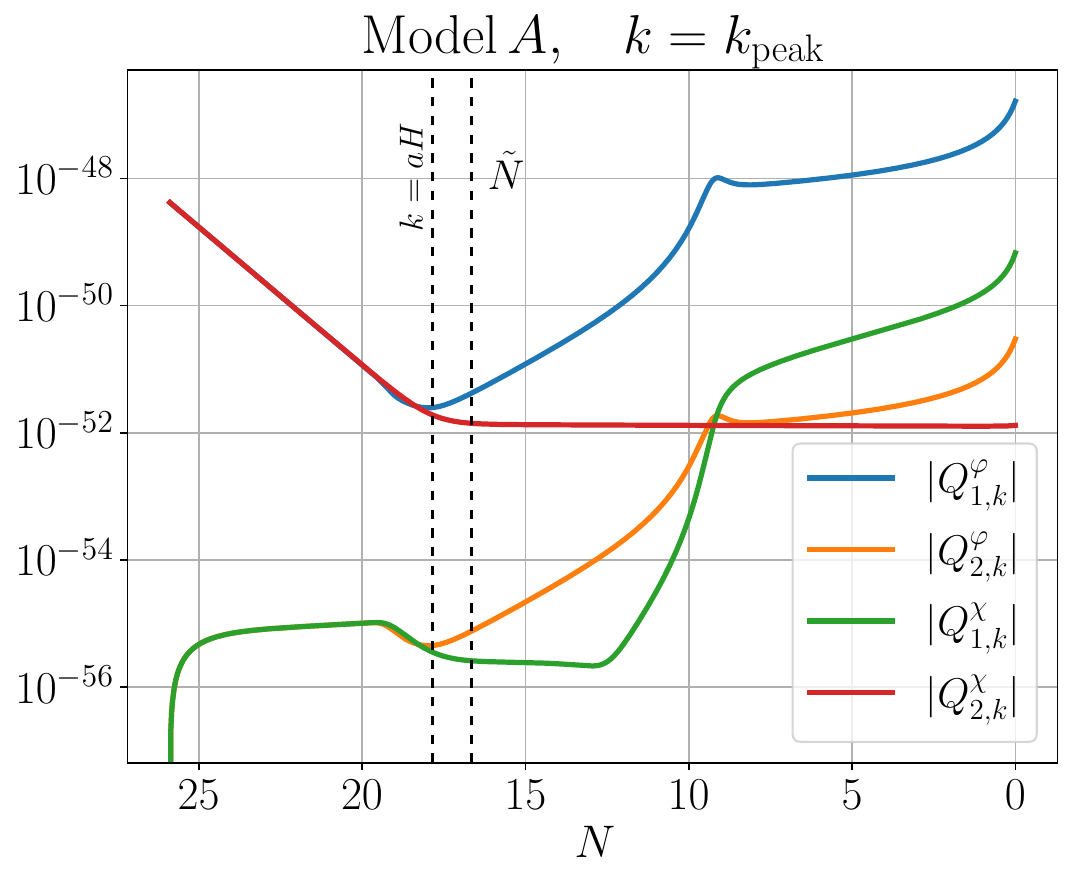}
    \picgap
    \includegraphics[width=\picwidth]{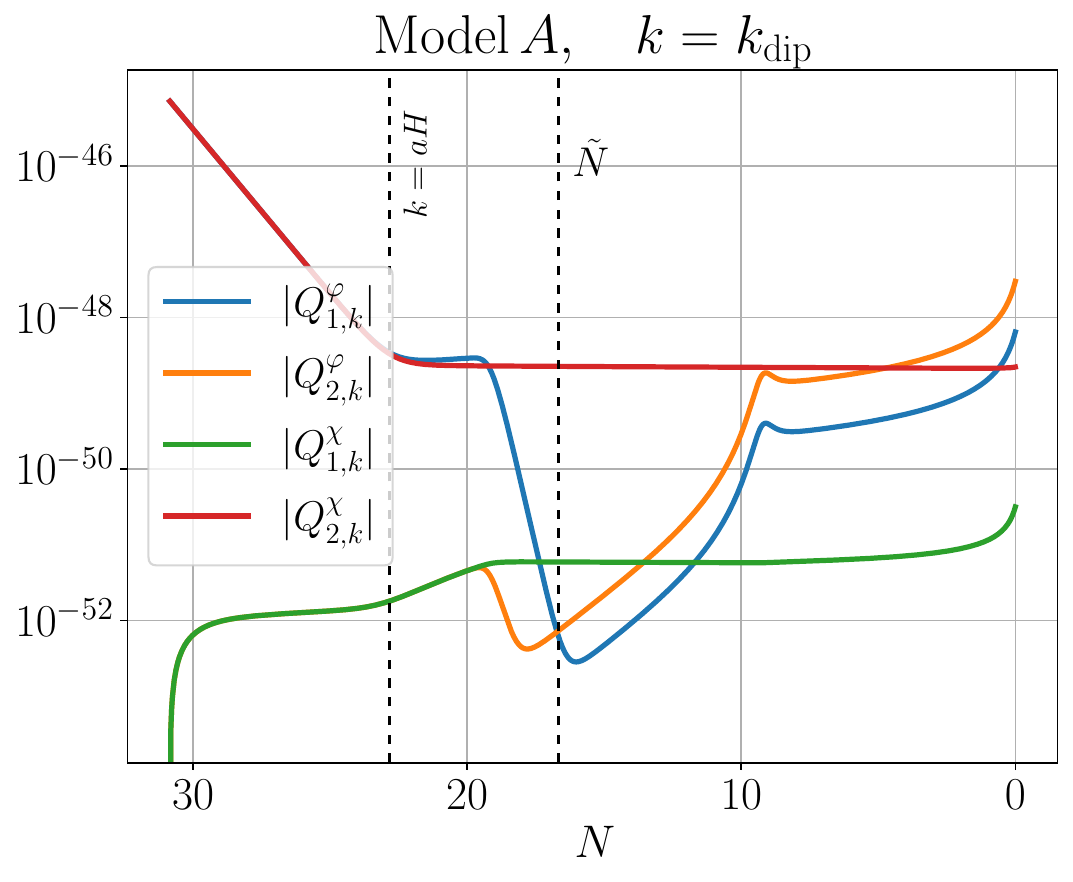}
    \includegraphics[width=\picwidth]{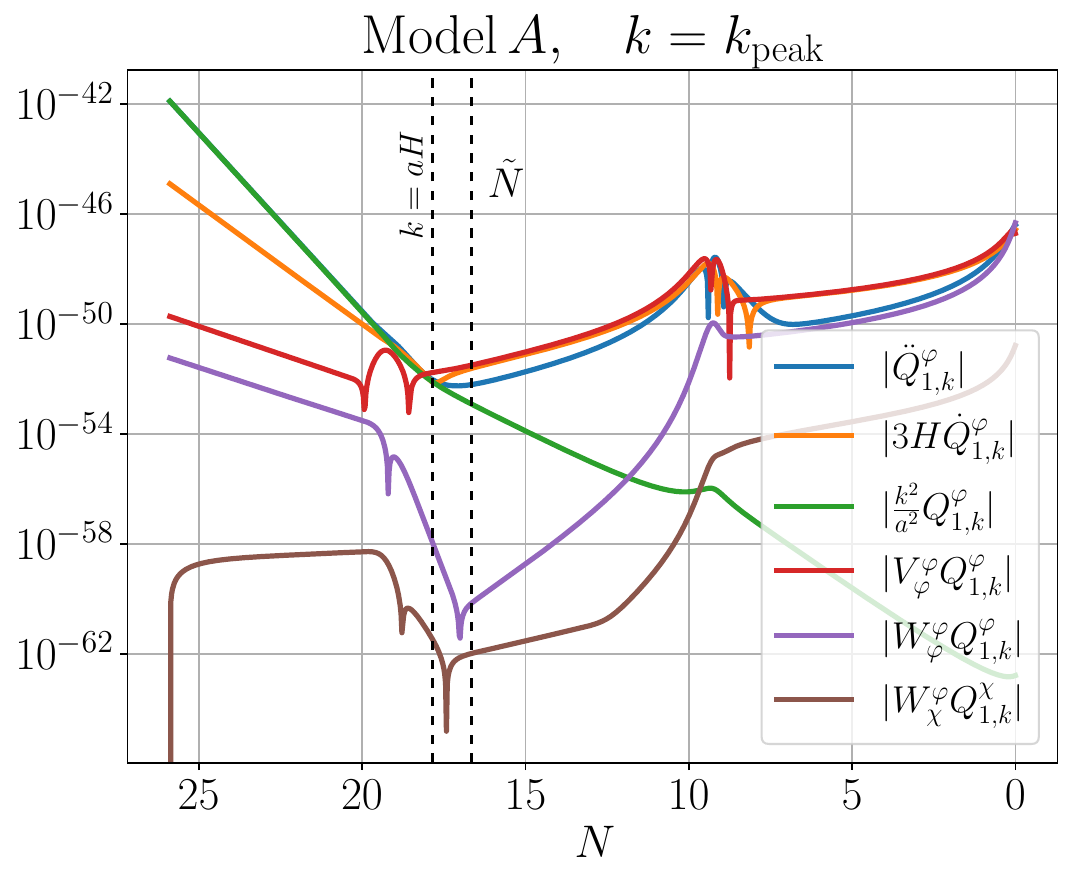}
    \picgap
    \includegraphics[width=\picwidth]{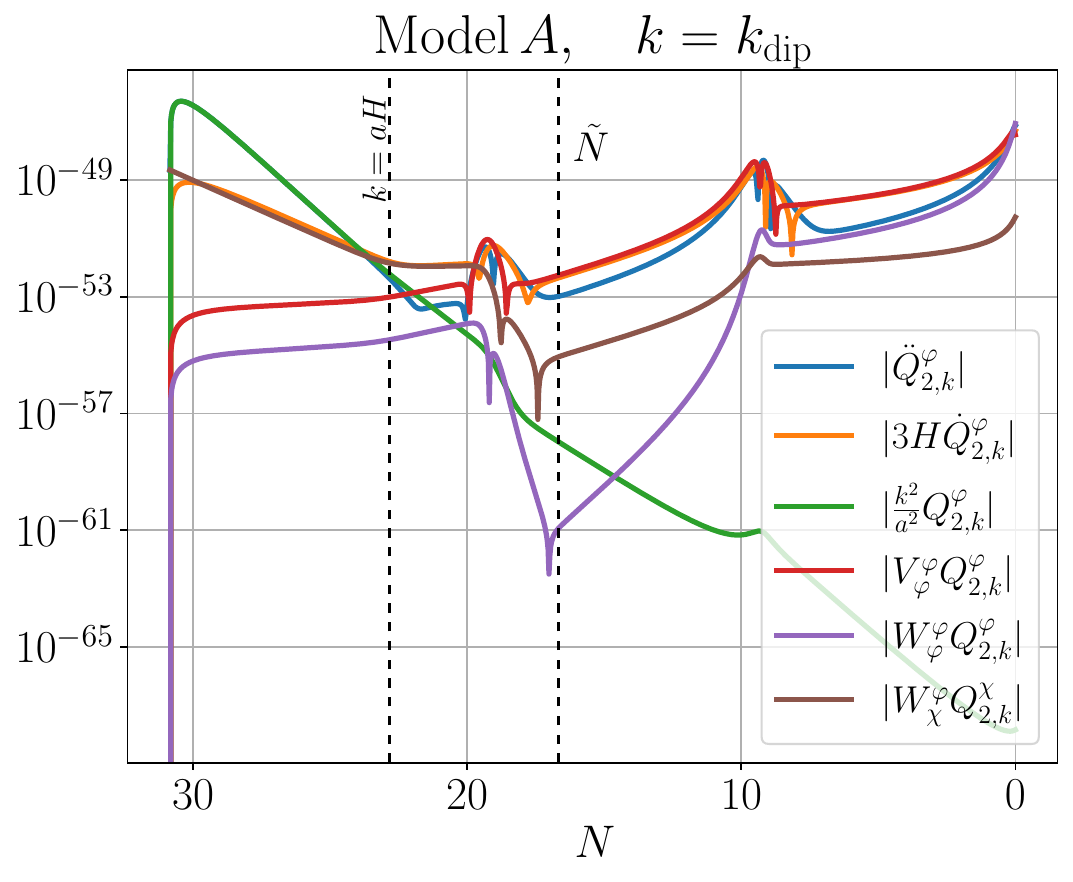}
    \caption{Top: evolution of the $Q^I_{i,k}$ perturbation components in Model $A$ for the maximum and minimum power spectrum modes, with $k_\text{peak} = 1.81\times 10^{15}$Mpc$^{-1}$ and $k_\text{dip} = 1.27\times 10^{13}$Mpc$^{-1}$.
    Bottom: different components of the evolution equation for the dominant modes.}
    \label{fig:modes_and_eoms}
\end{figure*}

The good agreement between both approximations and the full dynamics around $k_\text{peak}$, and the disagreement around $k_\text{dip}$, can be understood by studying the behavior of the modes. Fig.~\ref{fig:modes_and_eoms} depicts mode evolutions with the full dynamics in Model $A$.

In the top left panel of Fig.~\ref{fig:modes_and_eoms}, we show the time evolution of the absolute values of $Q^\varphi_{1,k}$, $Q^\varphi_{2,k}$, $Q^\chi_{1,k}$, and $Q^\chi_{2,k}$ for $k=k_\text{peak}$. We see that $Q^\varphi_{1,k}$ dominates over the rest at the end of inflation, providing the main contribution to the power spectrum. This is expected: $Q^\varphi_{1,k}$ is the mode that starts from the Bunch--Davies vacuum and grows due to the background dynamics. Its counterpart, $Q^\varphi_{2,k}$, also grows, but since it starts from zero it never reaches comparable values. The bottom left panel shows the different components of the $Q^\varphi_{1,k}$ equation. The $W_\chi^\varphi Q^\chi_{1,k}$ term remains subdominant throughout the entire evolution, explaining the goodness of the no-mixing approximations \eqref{eq:Qphidecoupled}, \eqref{eq:Qphidecoupled_2} and rendering the dynamics of $Q^\varphi_{1,k}$ effectively single field at $k=k_\text{peak}$.

The top right panel of Fig.~\ref{fig:modes_and_eoms} shows the mode evolutions for $k=k_\text{dip}$. This time, $Q^\varphi_{2,k}$ is the dominant mode at late times. The $Q^\varphi_{1,k}$ mode undergoes typical single-field behavior near a dip, becoming very small (mode mixing is again subdominant for its time evolution), allowing $Q^\varphi_{2,k}$ to take over. The bottom right panel shows that $W_\chi^\varphi Q^\chi_{2,k}$ is important for the evolution of $Q^\varphi_{2,k}$ around horizon crossing; this is due to the largeness of $Q^\chi_{2,k}$, which is much greater than $Q^\varphi_{2,k}$ at these times, as can be seen from the top panel (the same is true for the $Q^\varphi_{2,k}$ of the peak mode). Although the mixing term becomes negligible later, its contribution modifies the integrated evolution of $Q^\varphi_{2,k}$, enhancing its final value. As a result, the power spectrum dip in Fig.~\ref{fig:PR} is not as deep in the full dynamics as it is in the no-mix approximations without $Q^\varphi_{2,k}$. This is even clearer in Model $B$, which follows qualitatively similar behavior.

The dip is a generic feature of single-field USR models, as discussed in detail in \cite{Briaud:2025hra, Fujita:2025imc}. It can have observable consequences for CMB spectral distortions \cite{Ozsoy:2021pws} or the 21 cm signal \cite{Balaji:2022zur}. Even though the peaks are similar, the differences in the dip may help to observationally distinguish the two-field effective USR models from true single-field USR.

\section{\label{sec:conclusions}Conclusions}
In this paper, we studied inflationary models introduced in Refs.~\cite{Lorenzoni:2025gni, Lorenzoni:2025kwn}, consisting of an inflaton field $\varphi$ with an inflection point potential and a free, light spectator field $\chi$. We confirmed that, just like a single-field inflection point potential, the two-field setup leads to an enhanced curvature power spectrum.

Our results are identical to those presented in Refs.~\cite{Lorenzoni:2025gni, Lorenzoni:2025kwn}, but the interpretations differ. In Refs.~\cite{Lorenzoni:2025gni, Lorenzoni:2025kwn}, the authors studied the perturbations in a basis aligned with the field velocity, decomposing them into curvature and isocurvature components and emphasizing the two-field nature of the system. In contrast, we studied the perturbations in the inflaton-spectator basis and argued that the system is effectively single-field: the spectator slightly modifies the inflaton's background evolution, but the inflaton perturbations (responsible for curvature perturbations at late times) are effectively decoupled from the spectator perturbations. Since the spectator is practically frozen, its effect can be modeled to a good accuracy (within an order of magnitude in our example models) by simply lifting the inflaton potential by the constant spectator contribution $\frac{1}{2}m_\chi^2\tilde{\chi}^2$. This increases Hubble friction, decreasing the inflaton velocity and prolonging the USR-like phase, thus enhancing the perturbations.

Since the spectator is not totally frozen, the system doesn't enter into proper ultra-slow roll in the sense that the total second slow-roll parameter $\eta$ stays below $3$. However, the directed second slow-roll parameter $\eta_{\varphi}$ does exhibit USR behavior. We call this effective ultra-slow roll. The usual intuition and approximations related to single-field USR models mostly apply to effective USR. The only significant difference we observed was the tempering of the dip in the curvature power spectrum just below the peak: as the standard single-field mode decreases, a second, mixed mode becomes important, setting a lower bound on the power spectrum.

Refs.~\cite{Lorenzoni:2025gni, Lorenzoni:2025kwn} also presented alternative parameter values for the inflection point potentials \eqref{eq:V_PBH_A}, \eqref{eq:V_PBH_B} that, when solved as single-field models, mimic their two-field results. Our single-field construction, based on an additional constant $\frac{1}{2}m_\chi^2\tilde{\chi}^2$ term, is more straightforward and provides further insight into the role of spectator fields in inflection point models. As argued in Refs.~\cite{Lorenzoni:2025gni, Lorenzoni:2025kwn}, adding the spectator field can make the models more resilient against fine-tuning \cite{Cole:2023wyx, Stamou:2024lqf, Iovino:2025tcv, Profumo:2026qpn}, since order one changes in the scalar field parameters can compensate for much more severe tuning in other model parameters. Our work lets one treat the scalar field on an equal footing with the other parameters, making comparisons between them straightforward in principle. We leave detailed considerations of fine-tuning for future work.

While we have studied a free spectator field, our results are more general in spirit. Different spectator field potentials may yield different time evolutions for $\chi$, but as long as $\chi$ evolves slowly, we expect it to affect the curvature perturbation mainly through its contribution to $H$. This opens up a new way of modifying models with USR-like phases, by subtly manipulating the Hubble friction in regions where the inflaton velocity is extremely small.

\begin{acknowledgments}
SSL would like to thank the Indo-French Centre for the Promotion of Advanced Research (IFCPAR/CEFIPRA) for support of the proposal 6704-4 titled `Testing flavors of the early universe beyond vanilla models with cosmological observations’ under the Collaborative Scientific Research Programme. ET is supported by the ``Fonds de la Recherche Scientifique'' (FNRS) under the IISN grant number 4.4517.08.
\end{acknowledgments}

\appendix

\section{\label{app:bg}Two-field evolution in the turning basis}

In this Appendix, we derive background results for the turning basis of two-field inflation. As laid out in \eqref{eq:turning_basis}, we start by defining the orthonormal basis vectors
\begin{equation} \label{eq:turning_basis_app}
\begin{gathered}
    \hat\sigma^I \equiv \frac{\dot\phi^I}{\dot\sigma} \, , \quad \dot\sigma \equiv \sqrt{\dot\phi^I\dot\phi_I} \, , \quad
    \hat s^I \equiv \epsilon^{JI}\hat\sigma_J \, , \\
    \hat\sigma^I\hat\sigma_I = \hat s^I\hat s_I = 1 \, , \quad \hat \sigma^I \hat s_I = 0 \, .
\end{gathered}
\end{equation}
Any vector $v^I$ can be uniquely decomposed into components along $\hat\sigma^I$ and $\hat s^I$ as $v^I = v_\sigma \hat\sigma^I + v_s\hat s^I$, with $v_\sigma = v_I\hat\sigma^I$, $v_s = v_I\hat s^I$.

Taking the time derivative of the normalisation condition $\hat\sigma^I \hat\sigma_I = 1$ gives $\hat\sigma^I\partial_t\sigma_I = 0$, that is, $\partial_t\sigma_I$ only has a component in the $\hat s^I$ direction. This component is, by definition, the turn rate $\omega$; that is,
\begin{equation} \label{eq:turn_rate}
    \partial_t \hat\sigma^I = \omega \hat s^I \, , \quad
    \partial_t \hat s^I = -\omega \hat \sigma^I \, .
\end{equation}
The second equation follows from \eqref{eq:turning_basis_app} and $\epsilon^{JI}\epsilon_{KJ} = -\delta^I_K$. We can interpret $\omega$ as the time derivative of the angle $\theta$ between $\hat\sigma^I$ and (say) the $\varphi$-axis.

\subsection{Field equations and solving the turn rate}
We can decompose the first and second field derivatives as
\begin{align}
    \label{eq:phi_dot_decomposed}
    \dot\phi^I &= \dot\sigma \hat\sigma^I \, , \\
    \label{eq:phi_dot_dot_decomposed}
    \ddot{\phi}^I &= \ddot{\sigma}\hat\sigma^I + \dot{\sigma}\omega\hat s^I \, .
\end{align}
The first of these follows directly from \eqref{eq:turning_basis_app}, and the second one is its time derivative, using \eqref{eq:turn_rate}. We can also decompose the field derivative as $V_I = V_\sigma \hat\sigma^I + V_s \hat s^I$; plugging these into the field equations \eqref{eq:phi_eom}, we can read off the $\hat\sigma^I$ and $\hat s^I$ components:
\begin{equation} \label{eq:sigma_eom}
    \ddot\sigma + 3H\dot\sigma + V_\sigma = 0 \, , \qquad
    \dot\sigma\omega + V_s = 0 \, .
\end{equation}
The Friedmann equation \eqref{eq:H} simply becomes
\begin{equation} \label{eq:H_2}
    3H^2 = \frac{1}{2}\dot\sigma^2 + V \, .
\end{equation}
In these equations, $\sigma$ can be regarded as the length of the field space trajectory, increasing monotonically as the field rolls.

Equation \eqref{eq:sigma_eom} now gives
\begin{equation} \label{eq:turn_rate_computed}
    \omega = -\frac{V_s}{\dot\sigma} = \frac{\epsilon^{IJ}V_I\dot\phi_J}{\dot\sigma^2} =
    \frac{V_\varphi \dot{\chi} - V_\chi\dot{\varphi}}{\dot{\varphi}^2 + \dot{\chi}^2} \, ,
\end{equation}
the result we quoted in \eqref{eq:omega}.

In Appendix~\ref{app:perts}, we will also use the following result, straightforwardly derived from the above results:
\begin{equation} \label{eq:omega_dot}
    \dot\omega = -V_{s\sigma} + \omega\frac{V_\sigma}{\dot\sigma} - \omega\frac{\ddot\sigma}{\dot\sigma}
    = -V_{s\sigma} - \omega H \qty(3 + 2\epsilon - 2\eta) \, ,
\end{equation}
where $V_{s\sigma} \equiv V_{IJ}\hat s^I \hat \sigma^J$.

\subsection{Slow-roll parameters}
We can write the slow-roll parameters \eqref{eq:sr_parameters} in terms of $\dot\sigma$ and $\ddot\sigma$ as
\begin{equation} \label{eq:sr_parameters_in_sigma}
    \epsilon = \frac{\dot\sigma^2}{2H^2} \, , \quad
    \delta = 2\epsilon + 2\frac{\ddot\sigma}{H\dot\sigma} \, ,
    \quad
    \eta = \epsilon - \frac{\ddot\sigma}{H\dot\sigma} \, .
\end{equation}

Our convention for the slow-roll parameters follows that of Refs.~\cite{Lorenzoni:2025gni, Lorenzoni:2025kwn}. Other works often use the conventions $\epsilon_1 = \epsilon_H = \epsilon$ and $\epsilon_2 = \delta$. Another convention often used in single-field inflation is
\begin{equation} \label{eq:eta_H}
    \eta_H \equiv -\frac{\ddot\sigma}{H\dot\sigma} = \eta - \epsilon \, .
\end{equation}
During slow roll, when $3H\dot{\phi}^I \approx -V^I$ and $3H^2 \approx V$, one can show that
\begin{gather}
    \label{eq:epsilon_V}
    \epsilon \approx \frac{V^IV_I}{2V^2} \equiv \epsilon_V \, , \\
    \label{eq:eta_V}
    \eta \approx \frac{V^{IJ}}{V}\frac{\dot\phi_I\dot\phi_J}{\dot\phi^K\dot\phi_K}
    =\frac{V^{IJ}}{V} \hat\sigma_I \hat\sigma_J
    \equiv \frac{V_{\sigma\sigma}}{V}
    \equiv \eta_V \, .
\end{gather}
In the single-field limit, these coincide with the usual potential slow-roll parameters \cite{Liddle:1994dx}.

Equations \eqref{eq:eta_H}--\eqref{eq:eta_V} have analogous versions for the directed parameters defined in \eqref{eq:directed_sr_parameters_1}, \eqref{eq:directed_sr_parameters_2}:
\begin{gather}
    \label{eq:eta_H_I}
    \eta_{H}^I \equiv -\frac{\ddot\phi^I}{H\dot\phi^I} = \eta^I - \epsilon \, , \\
    \label{eq:epsilon_eta_V_I}
    \epsilon^I \approx \frac{V_I^2}{2V^2} \equiv \epsilon_{V}^I \, , \qquad
    \eta^I \approx \frac{V^{II}}{V} \equiv \eta_{V}^I \, ,
\end{gather}
where the approximations again apply in slow roll, and the indices are not summed over.

\section{\label{app:perts}Perturbations in two-field inflation}
In Section~\ref{sec:perts}, we presented two versions of the Fourier space mode equations, \eqref{eq:Q_eom} for $Q^I$ and \eqref{eq:R_eom}--\eqref{eq:S_eom} for $\R$ and $\Ss$ defined in \eqref{eq:R_S}. In this Appendix, we outline how to derive \eqref{eq:R_eom}--\eqref{eq:S_eom} from \eqref{eq:Q_eom}. Similarly to the background equations above, we need to decompose all terms in the $Q^I$ e.o.m. into components along $\hat\sigma^I$ and $\hat s^I$. We also need to write the background quantities in $V^I_J$ and $W^I_J$ in this basis, as functions of the $\sigma$ and $s$ derivatives of $V$, together with $\omega$ and the slow-roll parameters instead of time derivatives of $\phi^I$.

Let us start by writing
\begin{equation} \label{eq:Q_decomposed}
    Q_k^I = \frac{\dot{\sigma}}{H}\R_k \hat\sigma^I + \frac{\dot{\sigma}}{H}\Ss_k \hat s^I \, .
\end{equation}
Then
\begin{equation} \label{eq:Vdd_decomposed}
\begin{aligned}
    V^I_J Q_k^J
    &= \frac{\dot\sigma}{H}\qty(V_{\sigma\sigma} \R_k + V_{\sigma s} \Ss_k)\hat\sigma^I \\
    &+ \frac{\dot\sigma}{H}\qty(V_{s\sigma} \R_k + V_{ss} \Ss_k)\hat s^I \, .
\end{aligned}
\end{equation}
We further eliminate $V_{\sigma s}=V_{s\sigma}$ in favour of $\dot{\omega}$ and slow-roll parameters using \eqref{eq:omega_dot}.

We remind the reader that $\sigma$ and $s$ refer to the $\hat\sigma^I$ and $\hat s^I$ directions, and we do not sum over these indices even when they are repeated (we have moved all indices to the same height to indicate this).

For the $W^I_J$ term, we get
\begin{equation} \label{eq:W_decomposed}
\begin{aligned}
    W^I_J Q_k^J = &-\frac{1}{a^3}\frac{\dd}{\dd t}\qty(\frac{a^3}{H}\dot\sigma^2\hat\sigma^I\hat\sigma_J)Q_k^J \\
    \overset{\eqref{eq:turn_rate}}{=} &-\frac{\dot\sigma}{H}\qty(\frac{1}{a^3}\frac{\dd}{\dd t}\qty(\frac{a^3}{H}\dot\sigma^2)\R_k
    + 2\omega\epsilon H\Ss_k)\hat\sigma^I \\
    &-\frac{\dot\sigma}{H} \times 2\omega\epsilon H\R_k\hat s^I \, .
\end{aligned}
\end{equation}
The remaining time derivative has the standard single-field form, with
\begin{equation} \label{eq:Q_term_time_derivative}
    \frac{1}{a^3}\frac{\dd}{\dd t}\qty(\frac{a^3}{H}\dot\sigma^2)
    \overset{\eqref{eq:sr_parameters_in_sigma}}{=}2H^2\epsilon(3+\delta-\epsilon) \, .
\end{equation}

The derivative terms are the most cumbersome; starting from \eqref{eq:Q_decomposed} and using \eqref{eq:turn_rate} repeatedly, we get

\begin{widetext}
\begin{align}
    \label{eq:Qdot_decomposed}
    \dot Q_k^I
    &= \qty( \frac{1}{2}\delta\dot\sigma\R_k + \frac{\dot\sigma}{H}\dot\R_k - \frac{\dot\sigma}{H}\omega\Ss_k)\hat\sigma^I + \qty(\frac{1}{2}\delta\dot\sigma\Ss_k + \frac{\dot\sigma}{H}\dot\Ss_k + \frac{\dot\sigma}{H}\omega\R_k)\hat s^I \, , \\
    \label{eq:Qdotdot_decomposed}
    \ddot{Q}_k^{I}
    &=\frac{\dot{\sigma}}{H}\qty[\frac{H}{\dot{\sigma}}\frac{\text{d}(\delta\dot{\sigma}/2)}{\text{d}t}\R_k+\frac{H\delta}{2}\dot{\R}_k+\frac{H\delta}{2}\left( \dot\R_k - \omega\Ss_k\right)+\frac{\text{d}}{\text{d}t} \left( \dot\R_k - \omega\Ss_k\right)
    -\omega\qty(\frac{H\delta}{2}\Ss_k + \dot\Ss_k + \omega\R_k)]\hat{\sigma}^I\nonumber\\
    &\hspace{0.4mm}+\frac{\dot{\sigma}}{H}\qty[\frac{H}{\dot{\sigma}}\frac{\text{d}(\delta\dot{\sigma}/2)}{\text{d}t}\Ss_k+\frac{H\delta}{2}\dot{\Ss}_k+\frac{H\delta}{2}\left(\dot\Ss_k + \omega\R_k\right)+\frac{\text{d}}{\text{d}t} \left(\dot\Ss_k + \omega\R_k\right)
    +\omega\qty(\frac{H\delta}{2}\R_k + \dot\R_k -\omega\Ss_k)]\hat{s}^I \, .
\end{align}
\end{widetext}
Here we used
\begin{equation} \label{eq:delta_sigma_dot}
    \frac{\dd}{\dd t}\frac{\dot\sigma}{H}
    = \dot\sigma\epsilon + \frac{\ddot\sigma}{H}
    \overset{\eqref{eq:sr_parameters_in_sigma}}{=} \frac{1}{2}\delta\dot\sigma \, ,
\end{equation}
and, using the slow-roll parameter formulas \eqref{eq:sr_parameters}, \eqref{eq:sr_parameters_in_sigma} together with the $\sigma$ e.o.m. \eqref{eq:sigma_eom}, we further have
\begin{align}
    \label{eq:delta_sigma_dot_2}
    &\frac{1}{2}\delta\dot\sigma \overset{\eqref{eq:sigma_eom}}{=}
    (\epsilon - 3)\dot\sigma - \frac{V_\sigma}{H} \\
    \label{eq:delta_sigma_dot_dot}
    \implies &\frac{\dd}{\dd t}\qty(\frac{1}{2}\delta\dot\sigma) =
    \frac{\dot{\sigma}}{H}\qty{H^2\left[2\epsilon(\delta-\epsilon)+3\eta\right]+\omega^2 - V_{\sigma\sigma}} \, .
\end{align}
When computing the time derivative of $V_\sigma$, note that both the field value and the $\hat\sigma^I$ basis vector are time-dependent, leading to
\begin{equation} \label{eq:V_sigma_dot}
\begin{aligned}
    \frac{\dd}{\dd t} V_\sigma =
    \frac{\dd}{\dd t} V_I\hat\sigma^I
    &\overset{\eqref{eq:turn_rate}}{=}
    V_{IJ}\hat\sigma^I\dot\phi^J
    +\omega V_I \hat s^I \\
    &\overset{\eqref{eq:sigma_eom}}{=}
    \qty(V_{\sigma\sigma} - \omega^2)\dot\sigma \, ,
\end{aligned}
\end{equation}
which we used in \eqref{eq:delta_sigma_dot_dot}.

We now have all the results we need. In summary, they relate to the e.o.m. of $Q_k^I$ as follows:
\begin{equation} \label{eq:Q_eom_decomposed}
    \overset{\eqref{eq:Qdotdot_decomposed}}{\ddot{Q}_k^I} + 3H\overset{\eqref{eq:Qdot_decomposed}}{\dot{Q}_k^I} + \frac{k^2}{a^2}\overset{\eqref{eq:Q_decomposed}}{Q_k^I} + \overset{\eqref{eq:Vdd_decomposed}}{V^I_JQ_k^J} + \overset{\eqref{eq:W_decomposed}}{W^I_JQ_k^J} = 0 \, .
\end{equation}
Substituting in the further simplifications discussed below the equations, we find that for the $\hat\sigma^I$ component, all terms proportional to $H^2$ cancel out, as do those proportional to $V_{\sigma\sigma}$ and $\omega^2$. Simplifying the rest, we finally obtain the $\R_k$ e.o.m. \eqref{eq:R_eom}.

The $\hat\sigma^I$ and $\hat s^I$ terms are symmetric under switching $\sigma \leftrightarrow s$, $\R \leftrightarrow \Ss$, and $\omega \leftrightarrow -\omega$, except for the $V_{\sigma\sigma}$ in \eqref{eq:delta_sigma_dot_dot}, the $\dot\omega$ arising from \eqref{eq:Vdd_decomposed} and \eqref{eq:omega_dot}, and by the extra $\hat\sigma^I$ term in \eqref{eq:W_decomposed}. This asymmetry leads to a different set of cancellations for the $\hat s^I$ driection: the $\omega^2$ terms still cancel, and so do terms proportional to $\R_k$, but the rest combine to give the $\Ss_k$ e.o.m. \eqref{eq:S_eom}.

\section{\label{app:quantization}Quantization}
The inflaton $\varphi$ and spectator $\chi$ are quantum fields. We can promote their perturbations $Q^I$ to quantum operators $\hat Q^I$ by introducing two sets of ladder operators $\hat a_{i,\vb{k}}$, $i=1,2$, with commutation relations
\begin{equation} \label{eq:a_commutators}
    \comm{\hat a_{i,\vb{k}}}{\hat a^\dagger_{j,\vb{q}}} = \delta_{ij}\delta^{(3)}(\vb{k} - \vb{q}) \, .
\end{equation}
As we can see from the classical equation \eqref{eq:Q_eom}, the perturbations couple. At quantum level, this coupling can be achieved by letting both $\hat Q^\varphi$ and $\hat Q^\chi$ depend on both sets of ladder operators. We introduce a total of four sets of mode functions $Q^{I}_{i,k}$, $I=\varphi,\chi$; $i=1,2$, and write
\begin{equation} \label{eq:Q_operators}
\begin{aligned}
    \hat Q^I(\vb{x},t) = \int &\frac{\dd k^3}{(2\pi)^{3/2}} \Big[
    Q^I_{1,k}(t) \hat a_{1,\vb{k}} + Q^{I*}_{1,k}(t) \hat a^\dagger_{1,-\vb{k}} \\
    &+ Q^I_{2,k}(t) \hat a_{2,\vb{k}} + Q^{I*}_{2,k}(t) \hat a^\dagger_{2,-\vb{k}}
    \Big] e^{-i\vb{k}\cdot\vb{x}} \, .
\end{aligned}
\end{equation}
We demand that $Q^{\varphi}_{i,k}$ and $Q^{\chi}_{i,k}$ satisfy the equations \eqref{eq:Q_eom} separately for $i=1$ and $i=2$; in other words, modes with equal $i$ mix, but modes with different $i$ don't. Then, clearly, the operators $\hat Q^I$ satisfy the correct coupled equations of motion with the correct commutation relations (up to normalization arising from initial conditions, see below), while also describing the correct number of degrees of freedom.

At sub-Hubble scales, the perturbations should be in the Bunch--Davies vacuum state \cite{Birrell:1982ix}, where they decouple (the $k^2/a^2$ term in \eqref{eq:Q_eom} overwhelms the coupling terms). In this limit (and this limit only), we associate $\hat a_{1,\vb{k}}$ with $Q^\varphi$ and $\hat a_{2,\vb{k}}$ with $Q^\chi$, setting the corresponding mode functions to the Bunch--Davies values, while setting the others to zero\footnote{The factor $a$ in the denominators arises from the fact that $a Q^I$ is the correct canonical perturbation variable instead of $Q^I$; see, e.g., \cite{Birrell:1982ix}.}:
\begin{equation} \label{eq:Q_bunch_davies}
\begin{gathered}
    Q^\varphi_{1,k}(t_0) = \frac{1}{a(t_0)\sqrt{2k}} \, , \quad
    \dot Q^\varphi_{1,k}(t_0) = -\frac{ik}{a} Q^{\varphi}_{1,k}(t_0) \, , \\
    Q^\chi_{2,k}(t_0) = \frac{1}{a(t_0)\sqrt{2k}} \, , \quad
    \dot Q^\chi_{2,k}(t_0) = -\frac{ik}{a} Q^{\chi}_{2,k}(t_0) \, , \\
    Q^\chi_{1,k}(t_0) = \dot Q^\chi_{1,k}(t_0) =  Q^\varphi_{2,k}(t_0) = \dot Q^\varphi_{2,k}(t_0) = 0
\end{gathered}
\end{equation}
at an early time $t_0$ when the mode is far inside the Hubble radius, $a(t_0)H(t_0) \ll k$.

The curvature power spectrum is related to the expectation value of $\hat \R^2$ in the quantum vacuum state annihilated by $\hat a_{i,\vb{k}}$. As such, it gets contributions from all four $Q^I_{i}$ modes:
\begin{equation} \label{eq:power_spectrum}
\begin{aligned}
    \expval{\hat\R(x,t)^2}{0}
    &\equiv \expval{\qty( \frac{H}{\dot\sigma}\hat Q_I(x,t)\hat\sigma^I)^2}{0} \\
    &= \int \dd \ln k \underbrace{\frac{k^2}{2\pi^2}\sum_{i=1,2} \qty|\frac{H}{\dot\sigma}Q_{i,k}^I\hat\sigma_I|^2}_{\equiv \PR (k)}
\end{aligned}
\vspace{0.0cm} %This helps distribute things nicely at the end.
\end{equation}
yielding the result quoted in \eqref{eq:PR}. Note the two different meanings of the `hat' here: $\hat \R$ and $\hat Q^I$ are quantum operators, while $\hat\sigma^I$ is a unit basis vector in the field space.

\setlength{\parskip}{0pt} %Fix dangling final short paragraph.

The two-field quantization procedure described here was used earlier in \cite{Achucarro:2010da, Achucarro:2016fby}. The quantization can also be carried out in another basis, such as the $(\R, \Ss)$ one; predictions for physical quantities are basis-independent.

\bibliography{bibliography}

\end{document}